\documentstyle[prl,aps,epsf]{revtex}
\begin{document}
\twocolumn[\hsize\textwidth\columnwidth\hsize\csname@twocolumnfalse%
\endcsname
\draft
\preprint{SU-ITP \# 97/33}
\title{SO(5) Superconductors in a Zeeman Magnetic Field}
\author{Jiang-Ping Hu and Shou-Cheng Zhang}

\address{\it Department of Physics, McCullough Building, Stanford
  University, Stanford  CA~~94305-4045}
\maketitle
\begin{abstract}
\end{abstract}
{\sl The generic symmetry of a system under a uniform Zeeman
magnetic field is $U(1)\times U(1)$. However, we show that $SO(5)$
models in the presence of a finite chemical potential and a finite
Zeeman magnetic field can have a exact $SU(2)\times U(1)$
symmetry. This principle can be used to test $SO(5)$
symmetry at any doping level.}

\pacs{ PACS numbers: 74.20.Mn, 74.25.Ha, 71.10.-w }
]

A fundamental question one can ask in connection with high $T_c$
superconductors is whether they are in the same universality class
of conventional $d$ wave BCS superconductors. While many aspects
of high $T_c$ superconductors are anomalous and quantitatively
different from conventional BCS superconductors, no sharp distinction
based on symmetry has been made so far. In the absence of a
external magnetic field and spin anisotropy, the symmetry of the
Hamiltonian is $SU(2)\times U(1)$, where the $U(1)$ charge
symmetry is spontaneously broken in the superconducting state.

A notable exception is the idea of $SO(5)$ symmetry between
antiferromagnetism (AF) and superconductivity (SC)\cite{so5}. This
theory predicts a finite temperature bi-critical point with a
enlarged $SO(5)$ symmetry at the transition point between AF and
SC. It also predicts a spin triplet $\pi$ resonance\cite{demler}
in the SC state which can be interpreted as the pseudo Goldstone mode
associated with the spontaneous symmetry breaking. However, in the
presence of a finite chemical potential, the explicit symmetry of
the Hamiltonian is still a direct product of the spin $SU(2)$ and
the charge $U(1)$ symmetry, which is not different from that of a
conventional BCS system.

In this paper, we point out a remarkable symmetry property of
$SO(5)$ symmetric Hamiltonians. In the presence of a finite
chemical potential $\mu$ and a finite Zeeman magnetic field
$B$, the original $SO(5)$ symmetry is broken to $U(1)\times U(1)$.
Here the first $U(1)$ group describes the spin rotation symmetry
in a plane perpendicular to the applied magnetic field and the
second $U(1)$ group is the usual charge symmetry.
In fact, any generic spin invariant Hamiltonian in the presence
of a finite Zeeman field would have the same $U(1)\times U(1)$
symmetry.
From that point of view, $SO(5)$ symmetric models do not seem
to be different from any generic models once a chemical potential
or a magnetic field is applied. However, we will show that for
a special combination where $B=\mu$, the $SO(5)$ symmetric models
enjoy an enlarged $SU(2)\times U(1)$ symmetry, which is not shared
by generic models. Furthermore, this special $SU(2)\times U(1)$
symmetry at $B=\mu$ is equivalent to the original $SO(5)$ symmetry
in the absence of these fields. This gives a a powerful new tool
to test the $SO(5)$ symmetry {\it at any doping level}. The
original $SO(5)$ symmetry exists only at a particular doping
level where the AF to SC transition occurs. This point is very
difficult to reach in high $T_c$ superconductors because of
complicated doping chemistry, and has not yet been identified
experimentally. Under the new proposal, however, the $SO(5)$
symmetry can be revealed at any doping level, provided one
applies a Zeeman magnetic field. This new test can give a sharp
symmetry distinction between a $SO(5)$ superconductor and a
conventional BCS superconductor, and it can also distinguish
various explanations of the $\pi$ resonance.

Let us consider the following Hamiltonian
\begin{eqnarray}
 H = H_{SO(5)} - \mu Q - B S_z
\end{eqnarray}
where $H_{SO(5)}$ is a $SO(5)$ symmetric Hamiltonian which
commutes with the ten $SO(5)$ symmetry generators $L_{ab}$. (For
notations and definitions please see ref.\cite{so5}). $Q$ and
$S_z$ are members of $SO(5)$ symmetry generators $L_{ab}$, they
generate charge rotation and spin rotation in the $xy$ plane
perpendicular to the external Zeeman field. Since $SO(5)$ is a
rank two algebra, one can choose $Q$ and $S_z$ as the two mutually
commuting generators. For $B=0$, the generic symmetry of $H$ is
$SU(2)\times U(1)$, while for non-vanishing values of $B$, the
original $SO(5)$ symmetry of $H_{SO(5)}$ is broken explicitly to
the $U(1)\times U(1)$, generated by $Q$ and $S_z$.

However, $H$ has a exact enlarged symmetry $SU(2)\times U(1)$ at
$B=\mu$. At this point, both the chemical potential and the Zeeman
term can be combined as $-\mu Q_{\uparrow}$, where $Q_{\uparrow}$
and $Q_{\downarrow}$ measure the number of up spin and down spin
electrons respectively. Furthermore, we can define a $SU(2)$
subalgebra of the original $SO(5)$ algebra generated by
\begin{eqnarray}
\pi_{\downarrow} = \sum_{k}sgn(cosk_x-cosk_y)
c_{Q+k,\downarrow}c_{-k,\downarrow}, ~~\pi_{\downarrow}^+, ~~ Q_{\downarrow}.
\end{eqnarray}
It is easy to see that they form a closed $SU(2)$ algebra,
\begin{eqnarray}
J_1 &=& \frac{1}{2}(\pi +\pi^+), ~~J_2=\frac{i}{2}(\pi-\pi^+),
\nonumber \\
J_3 &=& \frac{1}{2}Q_{\downarrow}, ~~[J_{\alpha}, J_{\beta}]
 = i\epsilon_{\alpha\beta\gamma}J_{\gamma}
\label{su2}
\end{eqnarray}
Since the generators of this subalgebra are formed by linear
combinations of the original $SO(5)$ generators $L_{ab}$, they all
commute with $H_{SO(5)}$. Furthermore, since they only involve
down spin electrons, they commute with $-\mu Q_{\uparrow}$.
Therefore, we have proven that at $B=\mu$, $H$ has a $SU(2)\times
U(1)$ symmetry, generated by the $SU(2)$ algebra defined by
(\ref{su2}) and the $U(1)$ generator $Q_{\uparrow}$.

Mathematically, the new symmetry $SU(2)$ at $B=\mu$ is related
to the isomorphism between the $SO(5)$ and the $SP(4)$ Lie
algebras. The root diagrams of these two algebras can be obtained
from each other through a $45$ degree rotation. This exactly
corresponds to going from the $(Q,S_z)$ basis for the root diagram
to the $(Q_{\uparrow},Q_{\downarrow})$ basis. In the original
basis, the $-\mu Q$ breaks the $SO(5)$ symmetry into a
$SU(2)\times U(1)$ symmetry. In the new basis, it is then obvious
that the $-\mu Q_{\uparrow}$ should also break the $SP(4)$ symmetry
into a remaining $SU(2)\times U(1)$ symmetry.

Now we proceed to analyze the collective modes associated with this
new symmetry. For this purpose, it is useful to first see how the
new symmetry emerge in the Lagrangian formalism. The effective
Lagrangian with exact $SO(5)$ symmetry can be expressed as:
\begin{eqnarray}
{\cal L}= \chi (\partial_t n_a)^2 -\rho (\partial_k n_a)^2 -V(n),
\label{lagrangian}
\end{eqnarray}
where $V(n) = -\frac{\delta}{2} n^2_a+ \frac{W}{4}|n|^4$.
We can introduce a magnetic field and a chemical potential simultaneously
in the above Lagrangian by applying the following transformation:
\begin{eqnarray}
\partial_t n_{\alpha} &\longmapsto& \partial_t n_{\alpha}
-i\epsilon_{\alpha \beta \gamma} B_{\beta}n_{\gamma}, \alpha = 2, 3,
4;\nonumber \\
 \partial_t n_{i} &\longmapsto& \partial_t n_{i}-i\epsilon_{ij} \mu
n_{j}, i,j = 1,5.
\end{eqnarray}
Choosing  $\hat{B}= (0,0, -B)$, the Lagrangian becomes
\begin{eqnarray}
{\cal L} = &\chi& (\partial_t n_a)^2 -\rho(\partial_k n_a)^2 \nonumber
 \\
 &-&2i\chi(Bn_3
\partial_t n_2 - Bn_2
\partial_t n_3 - \mu n_1
\partial_t n_5 + \mu n_5
\partial_t n_1) \nonumber \\
&+&\chi [B^2(n_2^2+n_3^2)+\mu^2(n_1^2+n_5^2)] -V(n).
\end{eqnarray}
Denoting $\hat{M} = (n_1, n_2, n_5, n_3)$, and taking $B=\mu$, we can
rewrite the above equation into the following form:
\begin{eqnarray}
{\cal L} = &\chi& (\partial_t n_a)^2 -\rho(\partial_k n_a)^2 +2i\chi \mu \hat{M}R
\partial_t \hat{M}^T \nonumber \\
&+&\chi \mu^2\hat{M}^2 -V(n),
\label{M-Lagrangian}
\end{eqnarray}
where $R$ is a four dimensional matrix,
 \[ R =
\left( \begin{array} {cc}
 0 & I \\
-I & 0\\
\end{array} \right).\]
Now we discuss the symmetry of above Lagrangian.  Obviously, except
the third term in above equation, all other terms have a exact $SO(4)$
symmetry in the $\hat{M}$ space. However not all of rotation will
keep the invariance of the third term. If $\hat{O}$ denotes a rotation
matrix in the $\hat{M}$ space, then it must satisfy
\begin{eqnarray}
  \hat{O}^T \hat{O} = 1;~~ \hat{O}^T R\hat{O} = R.
\end{eqnarray}
in order to keep the Lagrangian (\ref{M-Lagrangian}) invariant.
Since $SO(4) \cong SU(2)\times SU(2)$, we immediately find one of the
$SU(2)$ subgroup whose  generators  are defined by  the following
matrix:
\begin{eqnarray}
G_{1} &=& \frac{1}{2} \left( \begin{array} {cc}
 \sigma_y&0\\
 0&\sigma_y  \\
\end{array} \right),~~ G_{2} =\frac{1}{2} \left(  \begin{array} {cc}
0& -i\sigma_{x}  \\
i\sigma_{x} & 0 \\
\end{array} \right),\nonumber \\
G_{3} &=&\frac{1}{2} \left( \begin{array} {cc}
 0&i\sigma_z\\
 -i\sigma_z&0 \\
\end{array} \right), ~~[G_{\alpha}, G_{\beta}] =
 i\epsilon_{\alpha\beta\gamma}G_{\gamma};
\end{eqnarray}
These matrices also have the following properties
$$ [G_{\alpha}, R] = 0.$$
Therefore,  $G_{\alpha}$, together with $R$, generates a symmetry
$SU(2)\times
U(1)$. The Lagrangain (\ref{M-Lagrangian}) is invariant
under above $SU(2)\times U(1)$ transformations.
By Noether's theorem, each internal symmetry is associated to a
conserved
charge.  From the infinitesimal variations of $\hat{M}$,
$$ \delta\hat{M}^T = iG_{\alpha}\hat{M}^T \delta \phi_{\alpha} +
 R\hat{M}\delta\phi_{R}, $$
we obtain the following conserved currents
\begin{eqnarray}
 j^R_t &=& 2\chi\partial_t \hat{M} R
 \hat{M}^T+2\chi\mu\hat{M}\hat{M}^T  \nonumber \\
 j^R_k &=&2\rho\partial_k \hat{M} R  \hat{M}^T  \nonumber \\
 j^{\alpha}_t &=&2i\chi\partial_t \hat{M} G_{\alpha}
 \hat{M}^T-2i\chi\mu\hat{M}RG_{\alpha}\hat{M}^T\nonumber \\
 j^{\alpha}_{k} &=&
 2i\rho\partial_k\hat{M}G_{\alpha}\hat{M}^T; \nonumber \\
 0 &=& \partial_t j^{R,\alpha}_t + \partial_k j^{R,\alpha}_k.
\end{eqnarray}
The associated conserved charges can be directly related to the
symmetry generators (\ref{su2}) in the Hamiltonian formalism:
\begin{eqnarray}
J_{\alpha} = \int dx j^{\alpha}_t\ \ ; \ \
Q_{\uparrow} = \int dx j^{R}_t.
\end{eqnarray}
Since the static potential
is explicitly broken from $SO(5)$ to
$SO(4)$, one might expect three massless Goldstone modes  and
one massive mode for this kind of symmetry broking.
However, there are two massless
modes and two massive modes in this case, because
the total Lagrangian (\ref{M-Lagrangian})
has lower $SU(2)\times U(1)$ symmetry than the
static potential.
We can pick one of the direction in $\hat{M}$ space and linearize the
mode equation around this direction, say $n_1$ ( superconductor
phase):
\begin{eqnarray}
\chi \partial_t^2 n_2 &=& \rho \partial_k^2n_2 - 2\mu\chi \partial_t
n_3 \nonumber \\
\chi \partial_t^2 n_3 &=& \rho \partial_k^2n_3 + 2\mu\chi \partial_t
n_2 \nonumber \\
\chi \partial_t^2 n_5 &=& \rho \partial_k^2n_5 \nonumber \\
\chi \partial_t^2 n_4 &=& \rho \partial_k^2n_4 - \chi\mu^2n_4
\end{eqnarray}
The last equation describes the massive modes with energy
$\omega_4=\mu$, which is associated with the explicit
symmetry breaking (from $SO(5)$ to $SO(4)$) of the static
potential. The third equation describes the usual Goldstone
massless mode( sound mode) of the superconductor with linear dispersion
$\omega_5=(\rho/\chi)k$. The first two
equations predict a new doublet-spin wave modes. One is massless, the
other is massive.  In the long wavelength limit, the energies of the
modes
are $\omega_2= v k^2, \omega_3 = 2\mu$. Therefore, there are always
two gapless modes, one with linear dispersion and the other with
quadratic dispersion, independent of the orientation of superspin.

It is also interesting to investigate the case where the $SO(5)$
symmetry is explicitly broken, but a projected $SO(5)$ symmetry
defined in Ref.\cite{project1} and Ref.\cite{project2}
is present. We can add a
term $-g(n_2^2+n_3^2+n_4^2)$ to the $SO(5)$ symmetric potential
$V(n)$, and choose $g>0$ so that AF is favored at half-filling
where $\mu=0$. In this case, the effective potential in the
presence of $B$ and $\mu$ is given by
\begin{eqnarray}
V_{eff}(n) &=& V(n)-g(n_2^2+n_3^2+n_4^2) \nonumber \\
&-& \chi[B^2(n_2^2+n_3^2)+\mu^2(n_1^2+n_5^2)]
\label{Veff}
\end{eqnarray}
For $B=0$, there is a AF to SC transition at
$\mu_c=\sqrt{g/\chi}$. For $\mu>\mu_c$, the system is in a
SC state. This SC state has a $\pi$ resonance mode
with frequency
\begin{eqnarray}
\omega_0=\sqrt{\mu^2-\mu_c^2}
\end{eqnarray}
A finite magnetic field $B$ causes a triplet Zeeman splitting of this
$\pi$ mode, where the lower mode vanishes at a critical value
\begin{eqnarray}
B_c=\sqrt{\mu^2-\mu_c^2}
\label{Bc}
\end{eqnarray}
of the Zeeman field. On the other hand,
from Eq. (\ref{Veff}), we see that a finite Zeeman magnetic field $B$
induces a SC to AF transition when $B$ exceeds the same critical
value $B_c$ as given by Eq. (\ref{Bc}). At $B=B_c$, the effective
potential $V_{eff}$ as given in Eq. (\ref{Veff}) is exactly
$SO(4)$ invariant in the $\hat{M}=(n_1,n_2,n_5,n_3)$ space. The
kinetic terms further break this symmetry to $SU(2)\times U(1)$.
Summarizing above discussions we conclude that {\it both exact and
projected $SO(5)$ symmetric models have a exact quantum
$SU(2)\times U(1)$ symmetry at a critical value of the Zeeman
magnetic field, which is the energy of the $\pi$ resonance mode
measured in the units of the magnetic field}.

From above discussions we see that there are only two remaining
massless modes at the $B=\mu$ point. It would be interesting to
formulate a low energy theory where the two other massive modes
are explicitly projected out. In the Lagrangian formalism, this
can be accomplished by dropping the $n_4$ degree of freedom, and discarding
the second time derivative terms in equation (\ref{M-Lagrangian}).
This corresponds to a effective low energy Hamiltonian of the
form:
\begin{eqnarray}
{\cal H}_{eff} = V(\hat M).
\label{Heff}
\end{eqnarray}
where $V(\hat M)$ is a $SO(4)$ symmetric potential which only depends
on the magnitude of the $\hat M$ vector. This Hamiltonian is to be
quantized by the following quantization condition:
\begin{eqnarray}
  [M^T,M]=\frac{i}{2} R.
\label{quantization}
\end{eqnarray}
This formulation gives us yet another way to understand the origin of
the $SU(2)\times U(1)$ symmetry. ${\cal H}_{eff}$ on a single site is
nothing but the Hamiltonian for a symmetric
two dimensional harmonic oscillator,
where $\hat M$ denotes the {\it phase space} coordinates of a two
dimensional harmonic oscillator, and the quantization condition
(\ref{quantization}) is nothing but the Heisenberg commutation
relation between the coordinates and momenta. A symmetric
two dimensional harmonic oscillator has more than the $SO(2)$ symmetry
of the coordinate space, but less than the $SO(4)$ symmetry of the
phase space. In fact, it has a $U(2)=SU(2)\times U(1)$ symmetry.
This discussion carried over straightforwardly to the case of
coupled oscillators with a global $U(2)=SU(2)\times U(1)$ symmetry.

The observation of the new $SU(2)\times U(1)$ symmetry gives us
the possibility of testing the $SO(5)$ symmetry of the original
model {\it at any doping}. Starting from a SC state at zero magnetic
field, the superspin lies in the $(n_1,n_5)$ plane.
Within the $SO(5)$ model, the only effect of a applied
Zeeman magnetic field is to split the $\pi$ triplet resonance mode.
The intensity and commensurability of each member of the triplet remain
the same. At a critical field $B_c$, there is a first order transition from
the SC state into the AF state where the superspin lies in the
$(n_2,n_3)$ plane. At the same time, one of the $\pi$ mode
softens to zero energy at $B_c$. {\it The exact coincidence of mode
softening transition and a first order transition is the signature
of the new symmetry}. As we shall see, in a generic system, either the
first order transition occurs before the mode softens to zero
energy, or the mode softening occurs before the first order
transition, in which case the system will have two separate
second order phase transitions.

All above discussions are based on the assumption where the original
model has a exact or projected $SO(5)$ symmetry. In order to
see the physical signature of the $SO(5)$ symmetry, it is useful
to study the effects of a finite chemical potential and Zeeman
magnetic field on models without $SO(5)$ symmetry.
A general Landau-Ginzburg potential a approximate
$SO(5)$ model in the presence of a finite Zeeman magnetic field $B$
and chemical potential $\mu$ can be expressed as
\begin{eqnarray}
 V
=&-&\frac{\delta_c}{2}x-\frac{\delta_s}{2}y -\frac{\delta}{2}z \nonumber
\\
&+&\frac{W_c}{4}x^2+\frac{W_s}{4}(y+z)^2+
\frac{W_0}{2}x(y+z)
\end{eqnarray}
where  $n_1^2+n_5^2 =x,  n_2^2 +n_3^2= y, n_4^2 =z $, $\delta_c =
2\chi_c \mu^2+\delta$ and $\delta_s =
2\chi_s B^2+\delta$. There are two kinds of generic phase diagrams
described by this effective potential. The first type of phase
diagram is realized for $W_0^2>W_cW_s$ and is
depicted in Fig. 1. In this case, the Zeeman magnetic
field induces a first order phase transition from the SC state
to the AF state at a critical value of the magnetic field $B_c$.
However, the $\pi$ mode is still massive at $B_c$, which clearly
distinguishes this from the $SO(5)$ symmetric case. The first order
line terminates at a bi-critical point $T_{bc}$, where all static
properties have a emergent $SO(4)$ symmetry and all dynamic
properties have a $SU(2)\times U(1)$ symmetry.
The second type of phase diagram is realized for $W_0^2<W_cW_s$,
it describes two second
order phase transitions, with a intervening mixed phase region where
both SC and AF orders coexist, as shown in Fig. 2.
The mixed region shrinks to zero at a finite
temperature tetra-critical point $T_{tc}$.
In the mixed phase, there are also two gapless modes and two massive modes.
However, there is a major difference for the gapless modes between exact
and approximate $SO(5)$ models. The two
gapless modes in this approximate $SO(5)$ model both have linear
dispersion in mixed phase. In an exact $SO(5)$ symmetry model, as what
we pointed out before, there is one gapless mode with quadratic
dispersion, leading to a system with infinite compressibility at
the transition point\cite{project1}.

In conclusion we have discovered a new symmetry of $SO(5)$ models in
the presence of a finite Zeeman magnetic field $B$ and chemical
potential $\mu$. At the special point $B_c=\mu$, the static potential
has a exact $SO(4)$ symmetry and the full Hamiltonian has a exact
$SU(2)\times U(1)$ symmetry. These considerations also generalize
to the projected $SO(5)$ model, where the critical magnetic field
is shifted to $B_c=\sqrt{\mu^2-\mu_c^2}$, as given by equation
(\ref{Bc}). This observation gives the possibility to experimentally
test the $SO(5)$ symmetry at any doping level. The Zeeman magnetic field
can be experimentally realized by applying a magnetic in the
two dimensional plane\cite{bourges}, so that the orbital effects
can be minimized. Below the critical value $B_c$,
our theory predicts that the Zeeman magnetic field
will only split the resonance energy, but not change the intensity of the
$\pi$ resonance mode. The $\pi$ mode should also remain commensurate
at momentum $(\pi,\pi)$.
The critical value of magnetic field needed for reaching the exact
$SU(2)\times U(1)$ symmetry point can also be
expressed as $B_c=\omega_0/g\mu_B$, where $\omega_0$ is the neutron
resonance energy, $g$ is the electronic $g$ factor, and $\mu_B$ is
the Bohr magneton. Unfortunately, this value exceeds $100T$ for all
high $T_c$ superconductors where neutron resonance has been discovered.
While it is not realistic to reach such a high magnetic field,
one could imagine starting from sufficiently
underdoped materials where the neutron resonance energy is much
lower, or one can perform the proposed experiments on other
materials\cite{organic} where the intrinsic energy scales are much lower.

We would like to thank Y. Bazaliy, S.A. Kivelson, D.H. Lee and C.
van Duin for stimulating discussions. This work is supported by
the NSF under grant numbers DMR-9814289 and DMR-9400372. J.P. H.
is supported by the Stanford Graduate Fellowship Program.


\begin{figure*}[h]
\centerline{\epsfysize=5cm
\epsfbox{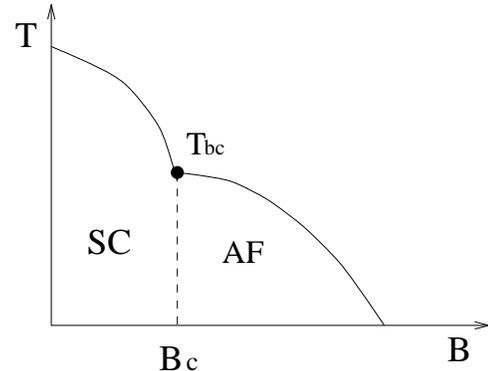}
}
\nonumber
\caption{Generic phase diagram of a approximate
$SO(5)$ superconductor in a Zeeman magnetic field.
The dashed line describes
a direct first order transition between SC and AF order.
At the bi-critical point $T_{bc}$, all static properties have
exact $SO(4)$ symmetry and all dynamic properties have exact
$SU(2)\times U(1)$ symmetry.
}
\label{fig1}
\end{figure*}

\begin{figure*}[h]
\centerline{\epsfysize=5cm
\epsfbox{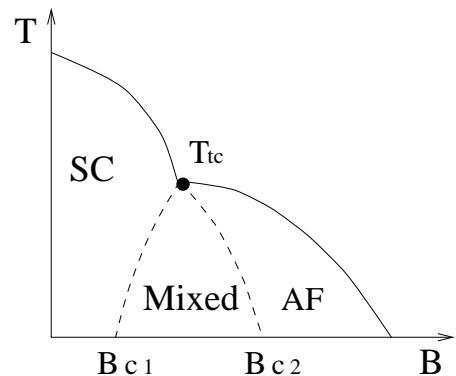}
}
\nonumber
\caption{Generic phase diagram of a approximate
$SO(5)$ superconductor in a Zeeman magnetic field.
The two dashed lines describe two second order phase
transitions, with a intervening mixed phase. The
two second order lines merge at a tetra-critical
point $T_{tc}$.}
\label{fig2}
\end{figure*}

\end{document}